\documentclass[final,authoryear,5p,times,twocolumn]{elsarticle}

\usepackage{amssymb}

\journal{Physics of the Earth and Planetary Interiors}

\usepackage{epsfig}
\usepackage{psfrag}
\usepackage{amsfonts}          
\usepackage{amsmath}           
\usepackage{ulem}
\usepackage{xspace}

\renewcommand{\vec}[1]{\mbox{\boldmath $ #1$}}

\providecommand{\Rs}{\ensuremath{R_s}\xspace}
\providecommand{\Rt}{\ensuremath{R_t}\xspace}
\providecommand{\Le}{\ensuremath{L}\xspace}
\renewcommand{\P} {\ensuremath{P}\xspace}

\providecommand{\etaa}{\ensuremath{\eta^{\ast}}\xspace}

\providecommand{\eg}{e.g.\xspace}
\providecommand{\A}{\textbf{aDD}\xspace}
\providecommand{\B}{\textbf{uTL}\xspace}
\providecommand{\C}{\textbf{sTL}\xspace}
\providecommand{\Done}{\ensuremath{I_1}\xspace}
\providecommand{\Dtwo}{\ensuremath{I_2}\xspace}

\usepackage{color}
\newcommand{\rev}[1]{#1}
\newcommand{\reva}[2]{#2}
\newcommand{\revb}[1]{}

\newcommand{\hide}[1]{}

\allowdisplaybreaks

\newlength{\shortwidth}
\newlength{\textwidtha}
\begin{document}

\begin{frontmatter}



\title{Double-diffusive convection in a rotating cylindrical annulus
  with conical caps}


\author{Radostin D.~Simitev}
\ead{Radostin.Simitev@glasgow.ac.uk}
\address{School of Mathematics and Statistics, University of
  Glasgow -- Glasgow G12 8QW, UK, EU}

\begin{abstract}
Double-diffusive convection driven by both thermal and compositional
buoyancy in a rotating cylindrical annulus with conical caps
is considered with the aim to establish whether a small fraction of
compositional buoyancy added to the thermal buoyancy (or vice versa)
can significantly reduce the critical Rayleigh number and amplify
convection in planetary cores. 
It is shown that the neutral surface describing the onset of
convection in the double-buoyancy case is essentially
different from that of the well-studied purely thermal case, and
does indeed allow the possibility of low-Rayleigh number convection. 
In particular, isolated islands of instability are formed by an
additional ``double-diffusive'' eigenmode in certain regions of the
parameter space.
However, the amplitude of such low-Rayleigh number convection is
relatively weak. 
At similar flow amplitudes purely compositional and double-diffusive
cases are characterized by a stronger time dependence compared to
purely thermal cases, and by a prograde mean zonal flow near the inner
cylindrical surface.   
Implications of the results for planetary core convection are briefly
discussed. 
\end{abstract}

\begin{keyword}
double-diffusive convection \sep 
buoyancy-driven instabilities \sep
planetary core 


\end{keyword}

\end{frontmatter}

\section{Introduction}

Convection in the cores of the Earth and the terrestrial planets is
of significant interest as it drives the dynamo processes that
generate and sustain the global magnetic fields of these bodies
\citep{Kono2002Recent,Jones2007}. 
Core convection is a double-diffusive process driven by density
variations due to non-uniform temperature and composition
\citep{BraginskyRoberts1995}. While ouble-diffusive phenomena are 
well-studied in oceanography, metallurgy, mantle convection and other
contexts \citep{Huppert1981,Turner1974,Turner1985,Schmitt1995}, their
manifestations in core convection remain poorly understood. 
It is thought that thermal and compositional buoyancy in
the Earth's core have comparable strength
\citep{Lister1995,Nimmo-2007}, and that temperature and concentration 
of light elements have widely different molecular diffusive time scales,
boundary conditions and source-sink distributions
\citep{BraginskyRoberts1995}. Yet, most planetary and geo-dynamo models 
consider only thermal convection or, at best, lump  
temperature and concentration into a single ``codensity'' variable.
The last approach is poorly justified, as it is only valid for equal
diffusivities and identical boundary conditions.  
\reva{q2}{Indeed, while eddy diffusivities due to small-scale
turbulent mixing tend to attain similar values, the turbulence in 
many cases, e.g.~weakly-convecting stratified layers, may not be as
fully developed to wipe out the large differences in molecular
diffusivities \citep{Busse2011}. At the same time, relatively small
variations in diffusivity ratios may have significant dynamical
effects \citep[e.g.][]{Simitev2005}.} 
So far, only few studies have been published where thermal and
compositional buoyancy are considered separately. \reva{m2}{\cite{Cardin1992}
performed an experimental investigation of thermochemical convection
in rotating spherical shell.} A double-diffusive numerical dynamo model
with a partly stable thermal gradient and  destabilizing compositional
component has been recently studied by \citet{Manglik2010}, as a
situation likely applicable to Mercury. Various driving scenarios
where thermal and compositional gradients 
are both destabilizing have been explored numerically by
\citet{Breuer2010}. All of these papers report significant differences
in their results to the single-diffusive (codensity) case and
emphasize the need for further investigation. 
The onset of double-diffusive convection in an axisymmetric
rotating system has been studied by \citet{Busse2002} in certain
asymptotic limits, and it was found that a small fraction of
compositional buoyancy could significantly reduce the critical
Rayleigh number, and thus amplify core convection. This prediction is
potentially very important, as it may shed light on  the 
thermodynamic state of the core and the  energy budget of the
geodynamo. However, concerted numerical simulations have so far failed
to confirm it \citep{Breuer2010}.     

With this motivation, the goals of this letter are to establish the
possibility of low-Rayleigh number double-diffusive convection, and to elucidate
the mechanisms by which thermal and compositional buoyancy interact. 
%
%
To this end, a simple model of a rotating cylindrical annulus with
conical end caps is considered here. This model has been very useful
in capturing the basic behaviour of nearly geostrophic convection in
the equatorial regions of planetary cores 
\citep{Busse2002review,Jones2007} and offers  
significant mathematical and computational advantages. The attention
is restricted here to the effects induced by the 
difference in diffusivity values, while the more realistic cases of
distinct 
boundary conditions and source-sink 
distributions are disregarded at present. The mathematical formulation and the methods of
solution are presented in Section \ref{sec1}. Sections \ref{sec2} and
\ref{sec3} describe linear and finite-amplitude properties of
double-diffusive convection. Conclusions and possible implications for
planetary cores are discussed in Section \ref{sec4}. 
\begin{figure}[t]
\psfrag{z}{\scriptsize $z$}
\psfrag{x}{\scriptsize $x$}
\psfrag{y}{\scriptsize $y$}
\psfrag{r}{\scriptsize $r_0$}
\psfrag{h}{\scriptsize $h$}
\psfrag{d}{\scriptsize $d$}
\psfrag{T1,C1}{\scriptsize $T_1,C_1$}
\psfrag{T2,C2}{\scriptsize $T_2,C_2$}
\psfrag{O}{\scriptsize $\vec{\Omega}$}
\psfrag{xi}{\scriptsize $\arctan \eta_0$}
\begin{center}
\vspace{-3mm}
\raisebox{0mm}{\epsfig{file=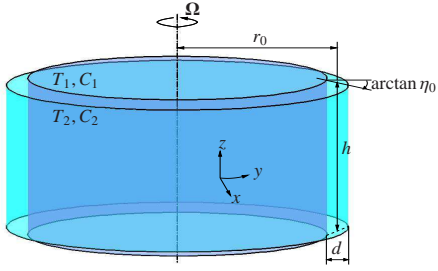,width=50mm,height=34mm,clip=}}
\end{center}
\vspace{-5mm}
\caption[]{Sketch of the rotating cylindrical annulus with conical
  end caps. \rev{Note that the sketch is not to scale with
  the limits of a small gap and a small angle of inclination of the
  conical caps.} \revb{q3} 
}
\label{f.010}
\end{figure}

\section{Formulation and methods of solution}
\label{sec1}
A cylindrical annulus with conical caps full of a two-component fluid,
and rotating about its axis of symmetry with an angular velocity $\Omega$
is considered. The configuration is shown in figure \ref{f.010}, and a
mathematical formulation of the problem given earlier by
\citet{Busse1986,Busse2002} is adopted. 
In particular, the inner and outer cylindrical walls are kept at constant
temperatures $T_0\mp \Delta T/2$, and at constant 
values of the concentration of the light element $C_0\mp \Delta C/2$, 
respectively, such that a density gradient opposite to the direction
of the centrifugal force is established as the basic state of the
system. \reva{q4}{The effect of the centrifugal force is similar to
  that of gravity in self-gravitating spheres and shells in that the
  buoyancy-driven motions occur in the same way as in the case when
  the force and the gradients of temperature and concentration are
  reversed. This formulation has the important advantage of being 
amenable to experimental realizations \citep[e.g.,][]{BusseCarrigan1974}.} 
The gap width $d$ of the annulus is used as a length scale, 
$d^2/\nu$ -- as the time scale, and $\nu \Delta T /\kappa$ and $\nu \Delta
C/\kappa$ -- as the scales of temperature and concentration of light
material, respectively. Here $\nu$ is the kinematic viscosity, and
$\kappa$ is the  thermal diffusivity.
A small-gap approximation, $d/r_0 \ll 1$, is assumed, where $r_0$ is 
the mean radius. This makes it possible to  neglect the spatial
variations of the centrifugal force, and of the temperature and concentration
gradients of the static state, and to introduce a Cartesian system of 
coordinates with the $x$-, $y$-, and $z$-coordinates in the radial,
azimuthal and axial directions, respectively. 
The Boussinesq approximation is adopted, in that the variation of
density,
\begin{gather}
\rho =\rho_0 \big(1-\gamma_t \Delta T (x- \Theta/\P) -\gamma_s \Delta C
(x- \Gamma/\Le)\big),
\end{gather}
is only taken into account in connection with the body forces acting
on the fluid. Here, $\gamma_t$ and $\gamma_s$ are the coefficients of
thermal and chemical expansion, and the other symbols are defined below.
\reva{q5}{The linear dependence on $x$ is unrealistic for the
concentration as it requires zero-concentration boundary conditions
\eqref{bcs}. A more realistic zero-flux condition would make the
problem rather involved \citep{BraginskyRoberts1995} 
and will divert from the main focus of this paper which is to
investigate the influence of diffusivities isolated from
the effects of boundary conditions. A discussion of different types of
boundary conditions related to core convection and the geodynamo can
be found, for example, in \citep{Kutzner2002From,Busse2006Parameter}.}
Assuming a small angle of inclination of the conical end caps  with 
respect to the equatorial plane, and taking into account that the
annulus is rotating, the velocity obeys approximately the
Proudman-Taylor theorem \reva{q3}{and can be described in first
  approximation by its geostrophic part}  
\begin{gather} 
\vec{u} = \nabla \times \vec{k}\psi(x,y,t) + \rev{O(\eta_0)},
\end{gather}
\rev{where $\eta_0\ll 1$ is the tangent of the said angle}.
Averaging over $z$, the governing equations for the leading order of the
dimensionless deviations of the  temperature $\Theta$, the
concentration $\Gamma$, and the stream function $\psi$ from the static
state of no flow can be written in the 2D cartesian form
\citep{Busse2002}  
\begin{gather}
\big((\partial_t-\nabla^2) + {\cal J}_\psi \big) \nabla^2\psi
-\etaa\partial_y\psi + \partial_y (\Rt\, \Theta+ \Rs\, \Gamma)
=0,\nonumber\\
\label{annulus}
\P \big(\partial_t+ {\cal J}_\psi \big) \Theta -\nabla^2\Theta
+\partial_y \psi=0, \\
\P  \big(\partial_t+ {\cal J}_\psi \big) \Gamma -\Le^{-1}\nabla^2 \Gamma
+\partial_y \psi=0,  \nonumber
\end{gather}
where  ${\cal J}_\psi=
(\partial_y\psi)\partial_x-(\partial_x\psi)\partial_y$, and the 
definitions of the rotation rate, Prandtl, Lewis, thermal and
compositional Rayleigh  numbers \etaa, \P, \Le, \Rt, and \Rs
are
\begin{gather}
\label{numbers}
\etaa = \frac{4\eta_0 \Omega d^3}{\rev{h} \nu}, \qquad 
\P=\frac{\nu}{\kappa}, \qquad \Le=\frac{\kappa}{D}, \\ 
\Rt=\frac{\gamma_t d^3 \rev{g} \Delta
  T}{\nu \kappa}, \qquad \Rs=\frac{\gamma_s d^3 \rev{g} \Delta
  C}{\nu \kappa}, \nonumber
\end{gather}
respectively.
Here, $D$ is the diffusivity of the light material, \reva{q6}{$h$ is
the axial length of the annulus}, and \reva{q4}{ $g=\Omega^2 r_0$ is
the average centrifugal acceleration analogous to gravitational
acceleration.}  
Fixed temperature and concentration, and stress-free BCs for the
velocity are assumed at $x=\pm 1/2$,
\begin{gather}
\label{bcs}
\psi=\partial_x^2\psi=\Theta=\Gamma=0 \quad \mbox{at} \quad x=1/2,
\end{gather}
while periodicity is imposed in the $y$-direction. 
For further details on the assumptions and for evidence of the utility
of this model to capture the dynamics of convection in rotating
spherical shells, the reader is referred to the reviews of
\citet{Busse2002review,Jones2007} and the 
references cited therein.
\begin{figure}[t]
\psfrag{Rtc}{$R_{t,\text{crit}}$}
\psfrag{ll}{$l$}
\begin{center}
\raisebox{0mm}{\epsfig{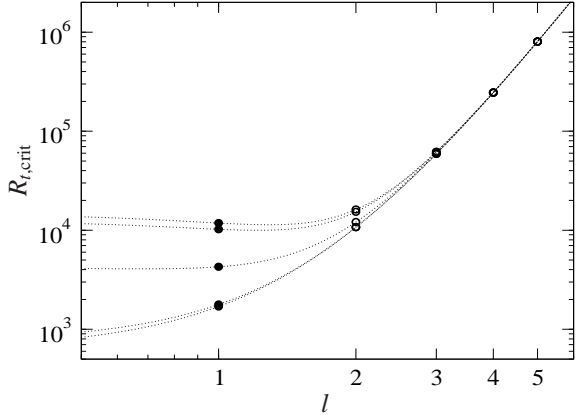}}
\end{center}
\vspace{-4mm}
\caption[]{\rev{The critical Rayleigh number $R_{t,\text{crit}}$
  of purely-thermal convection as a function of the radial wave number
  $l$ for $\alpha=5$, $\eta^\ast=600$, $\Rs=0$, and
  $P=10^{-2},10^{-1},1,10,10^2$ (from bottom to top). The solid
  circles indicate the preferred values of $l$.}  \revb{m1}
}
\label{f.020new}
\end{figure}

The linearized version of equations \eqref{annulus} allows an analytical
solution. The nonlinear equations \eqref{annulus} are solved
numerically by a modification of the Galerkin spectral method used
previously by \citet{OrBusse1987,SchnaubeltBusse1992}. The dependent variables
$\psi$, $\Theta$ and $\Gamma$ are expanded in functions satisfying the
boundary conditions     
\begin{gather}
\hspace{-12mm}
\begin{pmatrix} \psi \\ \Theta \\ \Gamma \end{pmatrix} =
\sum \limits_{l=0,n=1}^{\infty} 
\left[
\begin{pmatrix} \hat a_{ln} (t) \\ \hat b_{ln} (t) \\ \hat c_{ln} (t) \end{pmatrix}
\cos(n \beta y) +
\begin{pmatrix} \check a_{ln} (t) \\ \check b_{ln} (t) \\ \check  c_{ln} (t) \end{pmatrix}
\sin(n \beta y) \right] \times \\
\hspace{55mm}
\sin \big(l \pi \left(x+1/2 \right)\big). \nonumber 
\end{gather}
After projecting equations \eqref{annulus} onto the respective
expansion functions, a system of nonlinear ordinary differential
equations is obtained for the unknown coefficients $\hat a_{ln} (t)$,  
$\check a_{ln} (t)$, $\hat b_{ln} (t)$, $\check b_{ln} (t)$, $\hat
c_{ln} (t)$ and $\check c_{ln} (t)$. The system is integrated in time by a
combination of an Adams-Bashforth scheme for the nonlinear terms and a
Crank-Nicolson scheme for the diffusion and the other linear
terms. A truncation scheme must be introduced in practice: the
equations and the corresponding coefficients are neglected when
$l>N_x$ and $n>N_y$, where the truncation parameters $N_x$ and $N_y$
must be sufficiently large so that the physical properties of the
solution do not change significantly when their values are
increased. The computations reported in the following have been done
with $\beta=1$, $N_x=35$ and $N_y=55$.  

\section{The linear onset of double-diffusive convection}
\label{sec2}
Without loss of generality, small perturbations about the state of no
motion can be assumed to take the form    
\reva{m1}{}
\begin{gather}
\label{linansatz}
(\psi,\Theta,\Gamma)^\mathsf{T} = (\tilde{\psi},\tilde{\Theta},\tilde{\Gamma})^\mathsf{T}
\sin\big(\rev{l}\pi(x+1/2)\big)\,e^{i\alpha y+\lambda t}, 
\end{gather}
where $\alpha$ and $l$ denote the azimuthal \reva{m1}{and the radial} wave numbers, 
$\lambda = \sigma + i \omega$, with $\sigma \in \mathbb{R}$ and
$\omega \in \mathbb{R}$ being the growth rate and
the frequency of oscillations, respectively. The superscript~$^\mathsf{T}$ denotes transposition, and
$(\tilde{\psi},\tilde{\Theta},\tilde{\Gamma})^\mathsf{T}$ is a constant
component vector. Then, the linearised version of equations
\eqref{annulus} reduces to a matrix eigenvalue problem for $\lambda$ and 
$(\tilde{\psi},\tilde{\Theta},\tilde{\Gamma})^\mathsf{T}$, 
\begin{gather}
\label{eigenprob}
\begin{pmatrix} 
\displaystyle
-\frac{a^2+i\alpha\etaa}{a^2}& \displaystyle i\frac{\alpha \Rt}{a^2}&
\displaystyle i\frac{\alpha \Rs}{a^2}\\
\displaystyle-i\frac{\alpha}{\P}& \displaystyle-\frac{a^2}{\P}& 0\\
\displaystyle-i\frac{\alpha}{\P}& 0 &\displaystyle-\frac{a^2}{\P \Le}& \\
\end{pmatrix}
\begin{pmatrix} \tilde{\psi} \\ \tilde{\Theta} \\ \tilde{\Gamma} \end{pmatrix}
= \lambda \begin{pmatrix} \tilde{\psi} \\ \tilde{\Theta} \\ \tilde{\Gamma} \end{pmatrix},
\end{gather}
where $a^2 = \rev{l^2}\pi^2+\alpha^2$. \reva{m1}{In the rest of this
section attention is restricted to a single-roll convective
structure in radial direction by setting $l=1$. Equatorially-attached
``multicellular'' thermal convection has been previously found in
shells and annuli of finite gaps and convexly curved caps 
\citep[e.g.][]{Ardes1997,Plaut2005}. However, these geometries are
quite different from the small-gap limit considered here as they
provide radially inhomogeneous conditions for convection.
Figure \ref{f.020new} demonstrates that, in the small-gap limit and
for the parameter values discussed  below, $l=1$ is always the
preferred radial mode for the onset of thermal convection; the
nonlinear results of section \ref{sec3} further confirm the $l=1$ assumption. }  

\begin{figure*}[t]
\psfrag{Re(gam)}{$\sigma$}
\psfrag{Ra}{\Rt}
\psfrag{A}{\footnotesize{\A}}
\psfrag{B}{\footnotesize{\B}}
\psfrag{C}{\footnotesize{\C}}
\begin{center}
\hspace*{0mm}
\epsfig{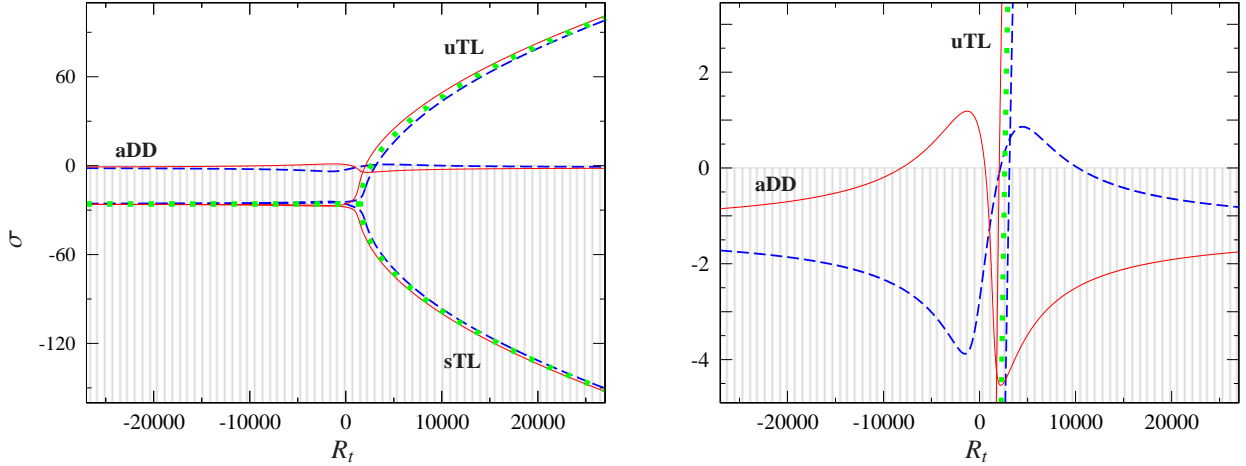}
\end{center}
\vspace{-4mm}
\caption[]{(Color online.) The growth rates,
  $\sigma=\text{Re}(\lambda)$, of the eigenmodes of   double-diffusive
  convection in the rotating annulus geometry as a   function of the
  thermal Rayleigh number \Rt at $\P=1$,   $\etaa=400$,
  $\alpha=4$, $\Le=20$,  $\Rs=500$ (solid red   line) of $\Rs=-500$
  (dashed blue line). The three possible modes are denoted by \A, \B,
  and \C in the example of $\Rs=500$.
  The basic state of no flow is linearly stable in the \rev{shaded
  region where the growth rate is negative}. Convection sets in the
  non-shaded  region.  The green dotted lines correspond to the well-studied purely thermal 
  modes of convection \eg \citep{Busse1986} at the same  parameter values
  (and $\Rs=0$, $L$ - arbitrary). The right panel is identical to the
  left one, only the scale of the $y$ axis is enlarged to show finer
  details. \revb{q7}
}
\label{f.020}
\end{figure*}

The solution to problem \eqref{eigenprob} can be found in analytical
form, and figure \ref{f.020} shows the growth rate of the perturbations,
$\sigma=\text{Re}(\lambda)$, as a function of the thermal Rayleigh
number \Rt for fixed values of the other parameters. The eigenmodes of 
purely thermal convection are also shown in the figure for comparison.
\begin{figure*}[htp]
\psfrag{Rt}{\Rt}
\psfrag{Rc}{\Rs}
\psfrag{alp}{$\alpha$}
\psfrag{eta}{\etaa}
\psfrag{P}{\P}
\begin{center}
\hspace*{1mm}
\epsfig{file=fig04.eps,width=\textwidtha,clip=}
\end{center}
\vspace{-4mm}
\caption[]{(Color online.) Neutral curves of double-diffusive
  convection in the rotating annulus geometry. Projections of the
  neutral surfaces onto
 (a) the $\alpha-\Rt$ plane,
 (b) the $\Rs-\Rt$ plane,
 (c) the $\etaa-\Rt$ plane, and
 (d) the $\P-\Rt$ plane.
  In all panels, the values of $\alpha=5$, $\P=10$, $\etaa=600$,
  $\Rs=-500$ (thick dashed blue lines) and $\Rs=500$ (thin solid red
  lines), and
  $\Le=$ 17 (innermost contour), 20, 30, 40 are kept fixed, except where
  they are given on the abscissa. As an example, the linearly unstable
  regions are shaded in the case $\Rs=500$,  $\Le=30$; the other curves
  form similar regions as well. The thick dotted green 
  lines (a single point in panel (b)) correspond to the well-known
  purely thermal Rossby wave modes of convection \eg \citep{Busse1986} at
  the same parameter values (and $\Rs=0$, \Le - arbitrary), and
  approximate closely the first asymptotic root \eqref{firstroot}. The black
  dash-dotted line represents the second asymptotic root \eqref{secondroot}.
}
\label{f.030}
\psfrag{absOM}{$|\omega|$}
\psfrag{Rc}{\Rs}
\psfrag{alp}{$\alpha$}
\psfrag{eta}{$\eta$}
\psfrag{P}{\P}
\begin{center}
\hspace*{0mm}
\epsfig{file=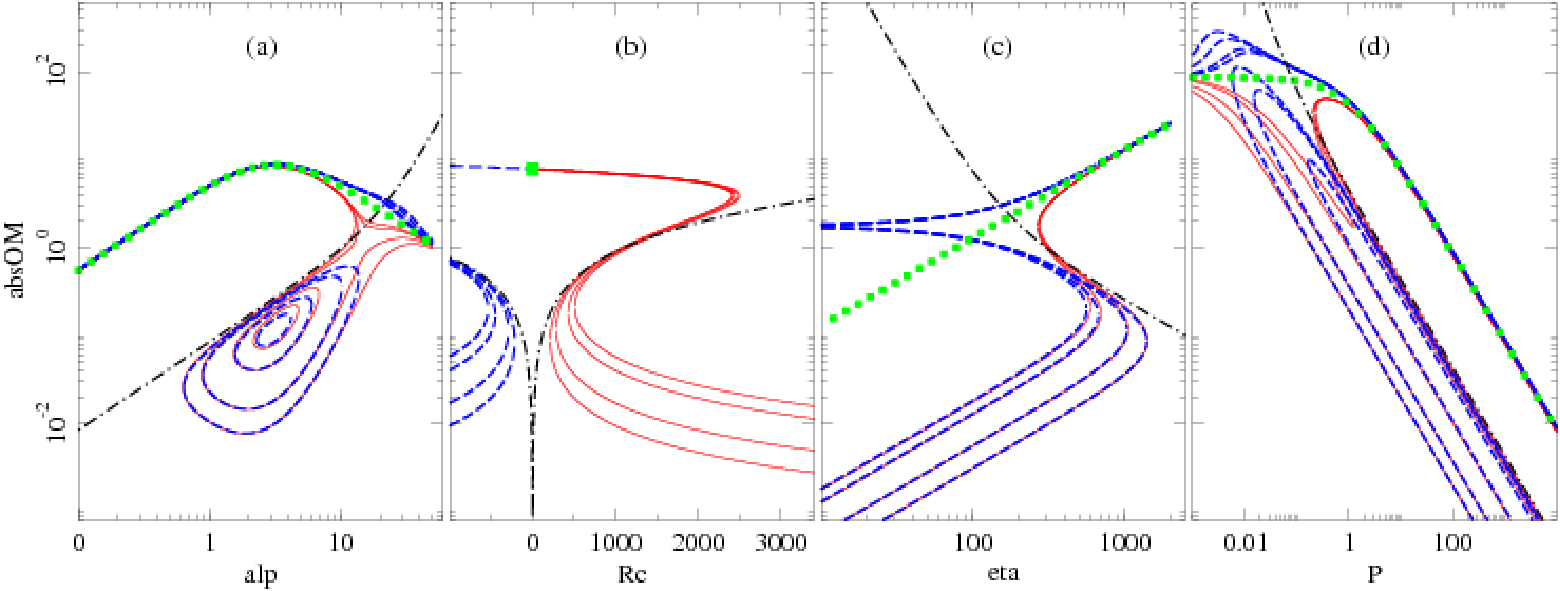,width=\textwidth,clip=}
\end{center}
\vspace{-4mm}
\caption[]{(Color online.) Amplitude of the frequency of oscillation,
 $|\omega|$, corresponding to the critical curves shown in figure
 \ref{f.030}. The line types and the parameter values are identical to
 those used in figure \ref{f.030}.
}
\label{f.040}
\end{figure*}
Because the matrix in \eqref{eigenprob} is of size $3\times3$, it can
have up to three distinct eigenmodes for typical parameter values. The
analogous eigenvalue problem for purely thermal  convection has a
matrix of size $2\times2$ that can have up to 2 eigenmodes at most.
Thus, a basic distinction between purely thermal and double-buoyancy
convection is the appearance of an additional ``double-diffusive''
eigenmode. The remaining two modes are analogous to the two possible
modes of purely thermal convection, as figure \ref{f.020} clearly
demonstrates. In figure \ref{f.020} and in the following, these three
possible modes are denoted by \A (additional Double-Diffusive 
mode), \B (unstable Thermal-Like mode), and \C (stable Thermal-Like
mode). The \A mode becomes unstable for smaller values of \Rt compared to
the \B mode. This provides a possibility for low-Rayleigh number
convection as suggested by \citet{Busse2002}. The growth rate of the \A
mode is a non monotonic function of \Rt, and it is remarkable that in
the case of a destabilizing compositional gradient  ($\Rs>0$), the \A
mode regains stability before the \B mode becomes unstable. This
limits the parameter space where low-Rayleigh number convection
occurs, and indicates the existence of isolated regions of instability.   
\hide{This observation provides a hint that the neutral (critical)
surfaces, i.e.~the surfaces in the parameter space that separate the
regions of linear stability ($\sigma<0$) from the regions of linear
instability ($\sigma >0$) of the system, may form isolated branches,
and that the usual scenario according to which the onset of convection
sets in only after a critical value of \Rt is exceeded may be
insufficient to describe the instability of the double-buoyancy system
under consideration.} 
\begin{figure*}[t]
\psfrag{At}{$\langle A \rangle_t$}
\psfrag{A}{$A$}
\psfrag{D1}{\Done}
\psfrag{D2}{\Dtwo}
\psfrag{Rs}{\Rt}
\psfrag{t}{$t$}
\begin{center}
\hspace*{0mm}
\epsfig{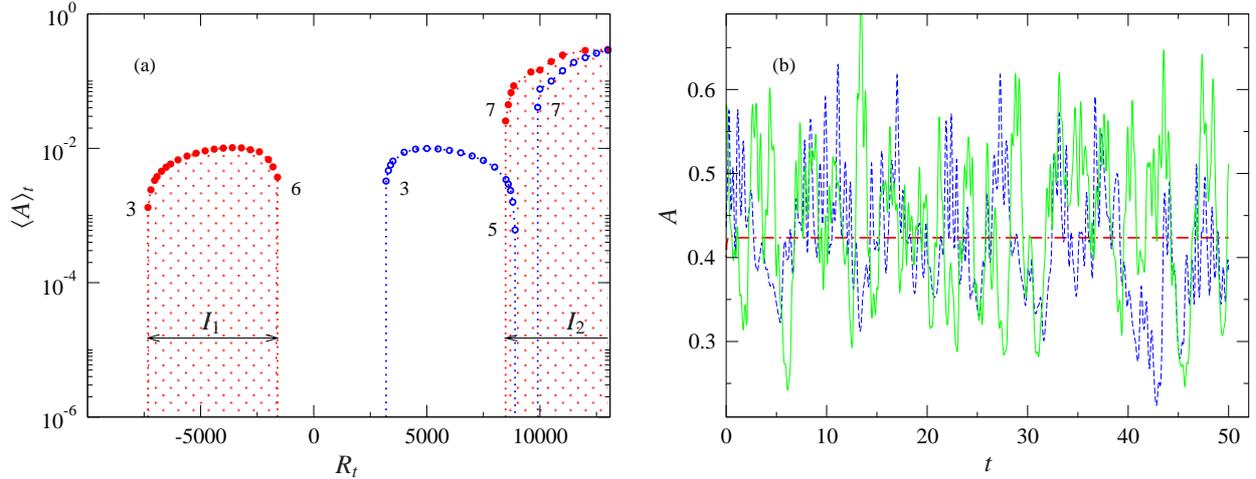}
\end{center}
\caption[]{(Color online.) (a) The time-averaged amplitude of
  convection $\langle A  \rangle_t$ as a function of the thermal Rayleigh number
  \Rt in the case $\P=10$, $\Le=20$,  $\etaa=600$, and $\Rs=500$
  (red) and $\Rs=-500$ (blue). In the  case $\Rs=500$, the region
  where convection occurs is shaded. For values of \Rt outside this
  region $A$  is a decaying function of \Rt, and the decay has been
  followed to values of $A$ smaller than $10^{-20}$ in all cases. The
  numbers shown near the onset of convection indicate the preferred
  wavenumber $\alpha$ in each case. (b) The amplitude of convection
  for $\P=10$, $\Le=20$, $\etaa=600$ and $\Rt=14050$, $\Rs=0$ (Case I,
  red dash-dotted line), $\Rt=0$, $\Rs=17000$ (Case II,
  blue dashed line) and $\Rt=\Rs=9100$ (Case III,
  green solid line).
}
\label{f.050}
\end{figure*}

The regions of linear stability ($\sigma<0$) and instability ($\sigma
>0$) in the parameter space are separated from each other by a neutral
surface. It is defined in implicit form by the characteristic equation
of the eigenvalue problem \eqref{eigenprob} where $\sigma=0$, 
\begin{gather}
\label{disprel}
\hspace{-10mm}
( i \omega \P +a^2) ( i \omega \P + a^2/\Le) \big((i \omega +a^2
)a^2 + i
\alpha \etaa \big)  \\
\hspace{10mm}
- \alpha^2 \Rt ( i \omega \P + a^2 / \Le) - \alpha^2 \Rs(i \omega \P
+ a^2 ) = 0. \nonumber
\end{gather}
Following \citet{Busse2002}, this equation is split into real and
imaginary parts from which the frequency of oscillations $\omega$ and
the critical value of any parameter of the problem as a function of
the remaining ones can be found in explicit analytical form.
Here, the thermal Rayleigh number \Rt is chosen as the principal
control parameter, because it offers the possibility of direct
comparison with the well-studied purely thermal case.
The five-dimensional neutral surface,
$\Rt=\Rt(\P,\etaa,\Le,\Rs,\alpha)$, is represented graphically by
its projections (neutral curves) onto the planes $\alpha-\Rt$,
$\Rs-\Rt$, $\etaa-\Rt$, and $\P-\Rt$ in panels (a, b, c, d) of figure
\ref{f.030} respectively, for fixed values of the remaining parameters
of the problem.   
The dependence on the Lewis number \Le is shown in the form of contour
lines in this figure thus exhausting all possible parameter
dependencies. The same approach is adopted to represent the
corresponding frequency $\omega=\omega(\P,\etaa,\Le,\Rs,\alpha)$ in
figure \ref{f.040}, where $|\omega|$ is plotted instead to
allow visualisation of finer details in the plots.  
The most prominent feature of the neutral curves is that they are 
multi-valued, and may split into closed, entirely isolated branches.
\reva{q8}{This can be understood from} the fact that the dispersion relation
\eqref{disprel} is a linear equation in \Rt, and a cubic equation in
$\omega$, so it has either one, two or three real roots as its
discriminant takes negative, zero and positive values when 
parameter values are continuously varied. The stability of the basic
state in the various regions formed thereby can be determined
from the sign of the growth rate $\sigma$ as described in
relation to figure \ref{f.020}. For example in the case $\Le=30$ of
figure \ref{f.030}, convection occurs within the regions that have
been shaded. 

The topology of the neutral curves of double-diffusive
convection is essentially different from that in the case of purely
thermal convection, also shown in figure \ref{f.030}. While in the
latter case for fixed values of the other parameters there is  
one and only one critical value of \Rt above which convection
occurs, in the former case up to three values of \Rt are needed to
specify stability criteria due to the multi-valued nature of the
neutral curves. Note that the critical wave numbers associated with
each of the three distinct critical values of \Rt are also different as
seen in figure \ref{f.030}(a). 
\reva{m3}{Occurrence of isolated regions of secondary instability has
  been reported in the case of quasi-geostrophic purely thermal
  convection by \cite{Plaut2002}. Note, that this is quite different
  from the isolated regions of primary instability discussed in this
  paper. It is likely that the double-diffusive case will exhibit even
  more complex behaviour in its transition to tertiary states, and
  this will be subject for future study.} 
Neutral curves with similar complex topology have been previously
reported in unrelated situations, e.g.~a differentially heated
inclined box \citep{Hart1971}, quiescent layers with density dependent
on two or more stratifying agencies with different diffusivities 
\citep{Pearlstein1981},  
isothermal shear flows 
\citep{Meseguer2002}, 
buoyancy-driven flows in an inclined layer 
\citep{Chen1989},
and penetrative convection in porous media 
\citep{Straughan1997}.

It is of interest to discuss the expressions 
\begin{gather}
\label{firstroot}
\Rt^{(1)}=\left[\frac{a^6}{\alpha^2}+\frac{1}{a^2}\left(\frac{\etaa
    \P}{1+\P}\right)^2\right]- \frac{a^2 \Rs^2 }{\eta^{\ast\,2} \P}-
    \frac{2\P \Rs}{1+\P}, \\
\omega^{(1)}=-\frac{\etaa \alpha}{a^2 (1+\P)} +
    \frac{a \Rs}{\etaa \P},\nonumber
\end{gather}
and
\begin{gather}
\label{secondroot}
\Rt^{(2)}=\frac{a^6}{\alpha^2} - \frac{a^2}{\eta^{\ast\,2} \P}\Rs^2,\\
\omega^{(2)}= -\frac{\alpha \Rs}{\etaa \P}\left[1+\frac{a^2 \Rs
    (1+\P)}{\eta^{\ast\,2} \P}\right], \nonumber
\end{gather}
derived by \citet{Busse2002} as solutions to the dispersion
relation \eqref{disprel} in the asymptotic limit of large \Le.
The first root corresponds to the well-studied thermal
Rossby waves, \citep[e.g.][]{Busse1986}, modified by the presence of
the second buoyancy component and describes the onset of the \B mode.
The physical nature of the second root (``the slow mode'') can be
understood from the observation that in the limit of large  \etaa the
second term in \eqref{secondroot} vanishes and the critical Rayleigh
number for the onset of Rayleigh-B\'enard convection in a non-rotating
plane layer is recovered \citep{Busse2002}. Thus, the additional
buoyancy provided by the compositional 
gradient, $\Rs \partial_y \Gamma$, counteracts the unbalanced
part of the convection-inhibiting Coriolis force, $\etaa \partial_y
\psi$, in equations \eqref{annulus}. 
Expressions \eqref{firstroot} and \eqref{secondroot} are shown in
figures \ref{f.030} and \ref{f.040}, and it can be seen that they
provide a good approximation to some pieces of the neutral curves even
for moderate and small values of \Le and \etaa. 
It has been implicitly assumed by \citet{Busse2002} that there is a
unique critical Rayleigh number above which convection sets in. This
led to the conclusion that the slow mode is the one preferred at onset, as
$\Rt^{(2)}$ is always smaller than $\Rt^{(1)}$.
The presented results show that this assumption is not always
correct, and that the multivalued nature of the neutral curves must be
taken into account. For example, when the concentration gradient is
destabilizing, 
$\Rt^{(2)}$ is, actually, the value at which convection decays as \Rt
is increased.

\section{Double-diffusive convection at finite amplitudes}
\label{sec3}

The linear results of section \ref{sec2} demonstrate that
low-Rayleigh number convection is indeed possible albeit the situation
is more complicated. Below, the question whether such low-Rayleigh
number flows are sufficiently vigorous to generate magnetic field is
addressed and finite-amplitude properties of double-diffusive
convection are 
explored. Finite-amplitude solutions are characterized by their mean
zonal flow, stream function, temperature and concentration
perturbations,  defined as
\begin{gather}
v_0(x,t)=\langle \partial_x\psi \rangle = \partial_x \Psi_0, \quad \Theta_0(x,t)=
\langle \Theta \rangle, \quad \Gamma_0(x,t)= \langle \Gamma \rangle,
\nonumber
\end{gather}
where 
\reva{q10}{}
$\langle f(y) \rangle =
L_y^{-1}\int_0^{\rev{L_y}} f(y) \textrm{d}y$ and $L_y=
2\pi/\beta$ is the basic periodicity length, and by the
amplitude of convection 
\begin{gather}
A^2= \sum \limits_{l=1,n=1}^{N_x,N_y} \left(\hat a_{ln}^2+\check
  a_{ln}^2\right).
\nonumber
\end{gather}
Figure \ref{f.050}(a) shows the time-averaged flow amplitude,
$\langle A \rangle_t$, of a sequence of cases with increasing value of
the thermal Rayleigh number \Rt and fixed values of the remaining
parameters. In full agreement with the linear theory, two regions of
convection are found, labeled \Done and \Dtwo in this figure. They
are separated by a region of vanishing flow. The amplitude of
convection in region \Done is more than an order of magnitude smaller
then that of the flow in region \Dtwo. Comparison with figure
\ref{f.020} indicates that the low-amplitude flow in \Done is
associated with the \A modes which are characterised by
relatively small values of $\sigma$, while the high-amplitude
convection in \Dtwo is likely associated with the \B modes. 
Because of its small amplitude, low-Rayleigh number double-diffusive
convection in region \Done is unlikely to be able to generate and
sustain magnetic fields on its own as will be further discussed below.
Within region \Done all computed solutions  are stationary, and for
this reason not illustrated, while  as \Rt is increased in region
\Dtwo a sequence of stationary, time-periodic, quasi-periodic and
chaotic solutions similar to those described in previous studies of
purely thermal convection, e.g.~\citep{HartBrummel},  is observed.

The additional physics introduced by the second buoyancy force makes
it difficult to compare directly double-diffusive convection  to the
much-better studied purely thermal case. A meaningful approach for
comparison is to consider cases with equally large
amplitudes. This is suggested by self-consistent MHD 
dynamo simulations where it has been established that sufficiently
vigorous turbulent flow is the primary condition for generation of
self-sustained magnetic fields \eg \citep{Simitev2005,Kutzner2002From}.     
For a comprehensive comparison the amplitude of the flow as a
function, for instance, of the thermal and compositional Rayleigh
numbers need to be computed. Then a contour plot of the data
$A(\Rs,\Rt)$ can be a useful comparison map as cases located on the
same energy level are expected to have similar ability for magnetic
field generation. However, the practical computation of such a surface has
proven too expensive even for the relatively simple annulus
model considered here. For this reason, the attention is restricted
below to a comparison of three representative cases: a purely thermal case, a
purely compositional case, and a mixed double-diffusive case,
henceforth Cases I, II and III, respectively.  The time-averaged amplitudes of
convection in Cases I, II and III are $\langle A \rangle_t = 0.42$,
0.42, 0.43, respectively. Although the values are not strictly equal,
additional simulations suggest that such small differences in amplitude are
not essential for the intended comparison. The three cases have
destabilizing thermal and compositional gradients, which is thought to
be appropriate for the Earth's core. Purposefully, the cases are
moderately rather than strongly driven to illustrate how simple known
properties are affected by the presence of a second buoyancy.
At these amplitudes the flows considered are associated with the \B
modes discussed previously, rather than with the newly-found \A
mode. This choice is justified as the \A modes do not produce
sufficiently vigorous flows with interesting structure, as already
discussed in relation to figure 
\ref{f.050}(a).  
\begin{figure*}[t]
\psfrag{ 0.4}{\hspace*{-2mm}\footnotesize 0.4}
\psfrag{ 0.2}{\hspace*{-2mm}\footnotesize 0.2}
\psfrag{ 0}{}
\psfrag{-0.4}{\hspace*{-2mm}\footnotesize -0.4}
\psfrag{-0.2}{\hspace*{-2mm}\footnotesize -0.2}
\psfrag{ 1}{\hspace*{-0.5mm}\raisebox{2mm}{\footnotesize 1}}
\psfrag{ 2}{\hspace*{-0.5mm}\raisebox{2mm}{\footnotesize 2}}
\psfrag{ 3}{\hspace*{-0.5mm}\raisebox{2mm}{\footnotesize 3}}
\psfrag{ 4}{\hspace*{-0.5mm}\raisebox{2mm}{\footnotesize 4}}
\psfrag{ 5}{\hspace*{-0.5mm}\raisebox{2mm}{\footnotesize 5}}
\psfrag{ 6}{\hspace*{-0.5mm}\raisebox{2mm}{\footnotesize 6}}
\begin{center}
\hspace*{0mm}
\epsfig{file=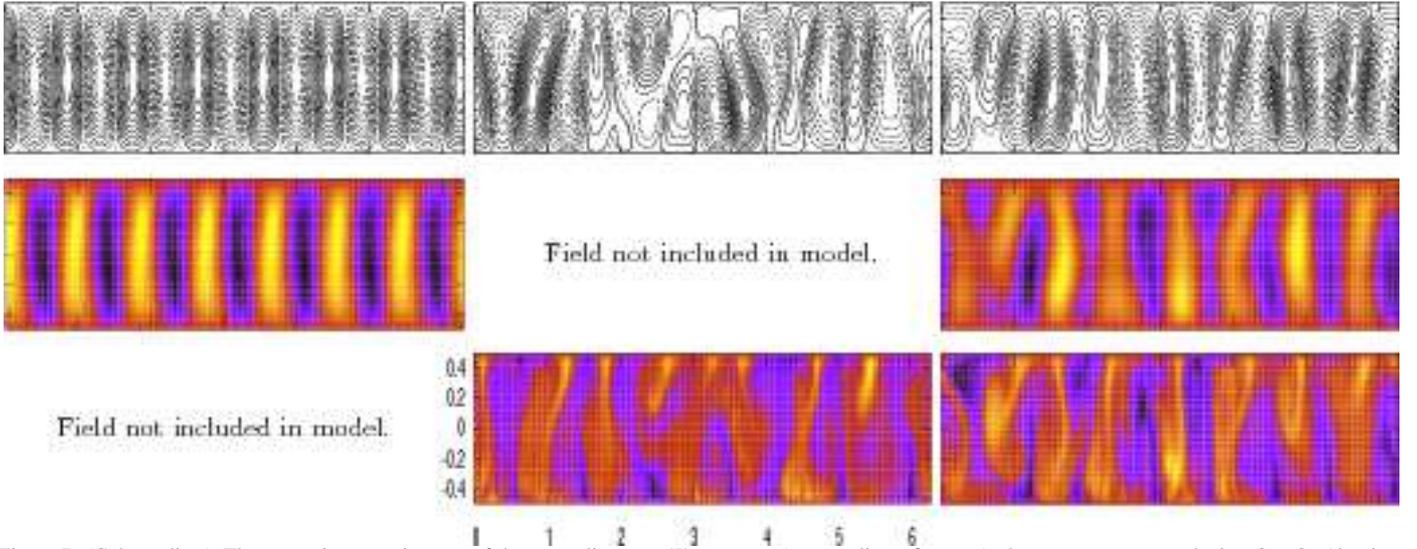,width=\textwidth,clip=}
\end{center}
\vspace*{-8mm}
\caption[]{(Color online.) The non-axisymmetric parts of the
  streamlines $\psi-\Psi_0=$ const.~(contour lines, first row), the
  \rev{temperature perturbation $\Theta-\Theta_0$} (density plot, second row)
  and the \rev{concentration perturbation $\Gamma-\Gamma_0$} (density plot,
  third row). The first, second and third columns correspond to the
  purely thermal Case I, to the purely compositional Case II, and to
  the double diffusive Case III, described in the caption of figure
  \ref{f.050}(b). \revb{q11}}
\label{f.060}
\end{figure*}

\begin{figure*}[t]
\psfrag{x}{$x$}
\psfrag{MF}{$\langle v_0\rangle_t$}
\psfrag{TP}{$\langle\Theta_0\rangle_t$}
\psfrag{CP}{$\langle\Gamma_0\rangle_t$}
\psfrag{RS}{$\langle \overline{vu}\rangle_t$}
\begin{center}
\hspace*{0mm}
\epsfig{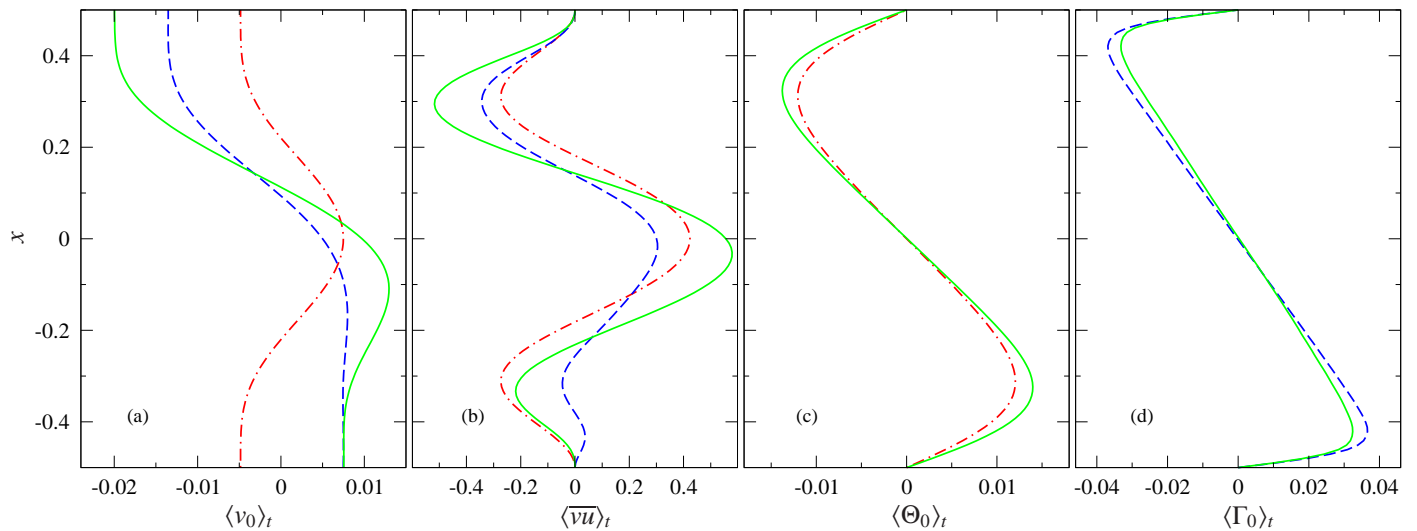}
\end{center}
\caption[]{(Color online.) Profiles of the time-averaged (a) mean
  velocity $\langle v_0\rangle_t$, (b) Reynolds
  stress $\langle \overline{vu}\rangle_t$, (c) mean temperature perturbation
  $\langle\Theta_0\rangle_t$, and (d) the mean concentration perturbation
  $\langle\Gamma_0\rangle_t$. Red dash-dotted lines indicate Case I, blue dashed lines
  indicate Case II, and green solid lines indicate Case III, described
  in the caption of figure  \ref{f.050}(b).
}
\label{f.070}
\end{figure*}

Figure \ref{f.050}(b) demonstrates that for comparable time-averaged
amplitude, the purely compositional Case II and the mixed Case III
have a highly chaotic time dependence while the purely thermal Case I
is stationary.  
The spatial properties of convection are shown in figure \ref{f.060}
where the streamlines of the flow are plotted for the
three cases along with the fluctuating parts of the temperature
perturbation $\Theta-\Theta_0$ and the concentration perturbation 
$\Gamma-\Gamma_0$. The plots represent snapshots at fixed moments in
time but they have been found to be representative for the three
cases. The purely thermal Case I shows a regular roll-like pattern which
does not change in time, while the structures corresponding to Cases II
and III are irregular and no periodic behaviour of the patterns in
time can be detected. 
\reva{m4}{These differences may be explained by the fact that Case I with}
\rev{ratio $\Rt/R_{t\text{,crit}}=1.52$ is far less supercritical than
Case II where $\Rs/R_{s\text{,crit}}=17.84$.}  
The predominant wave number of convection appears to be the same in
all three Cases, and remains equal to 7 
throughout the simulations. In comparison with the temperature
perturbation that shows relatively broad roll structures, the
concentration perturbation forms thinner plume-like structures,
consistent with the smaller compositional diffusivity.
%
The time- and azimuthaly-averaged properties of convection in the
three cases are compared in figure \ref{f.070}. The most obvious
difference is observed in the profiles of the time-averaged mean zonal
flow \reva{q13}{and the Reynolds stress. These
quantities are, indeed, related in that the mean flow is generated
primarily by the Reynolds stress \citep[e.g.~][]{Plaut2008,Busse2002}}. 
While in the purely thermal Case I the mean flow is symmetric with
respect to the mid-channel $x=0$, and retrograde at its ends
$x=\pm1/2$, in the purely compositional Case II it is asymmetric with
respect to $x=0$, retrograde at $x=1/2$ and prograde at $x=-1/2$. 
This asymmetry can be explained by the property that, unlike in the purely
thermal case, the value of \Rs in the compositional case is beyond the
onset of the mean-flow instability \citep{OrBusse1987}. 
The mean flow in the mixed Case III appears similar to the purely
compositional  case. The remaining panels in figure \ref{f.070} show
that the mean properties of the mixed case are similar to the
corresponding ones of the pure cases. In summary, it appears that
double diffusive convection associated with the \B modes can be
understood on the basis of the corresponding single-diffusive cases,
\reva{m2}{and that purely-thermal convection is more efficient in
imprinting its properties on the overall flow even when less
supercritical. This conclusion is confirmed by the experimental
results of \cite{Cardin1992} who studied thermochemical convection in
rotating spherical shells and found that the structure of
thermochemical flows is more like that of purely thermal convection. }

\section{Conclusion}
\label{sec4}
Convection driven by density variations due to differences in
temperature and concentration diffusing at different rates in a
rotating cylindrical annulus with conical end caps has been studied.  
It is shown by a linear analysis that the neutral surface
describing the onset of convection in this case has an essentially
different topology from that of the well-studied purely thermal
case. In particular, due to an additional ``double-diffusive'' 
eigenmode (\A), neutral curves are typically multi-valued and form
regions of instability in the parameter space which may be
entirely disconnected from each other. It is confirmed that the
asymptotic expressions for the critical Rayleigh number and frequency
derived by \citet{Busse2002} describe the onset of convection over an 
extended range of non-asymptotic parameter values but do not capture
the full complexity of the neutral curves. The results necessitate a revision of the assumption that there is a
unique critical value of the control parameter, \eg~\Rt, and call for
a better appreciation of the multivalued nature of the critical curves.
It is been found that finite-amplitude low-Rayleigh number
convection due to \A modes is possible over a wide parameter range.
However, the resulting flow amplitudes are significantly lower than
those of due to the familiar \B modes of convection. For this reason,
low-Rayleigh number flows are unlikely to be able to generate and
sustain magnetic fields on their own.   
In order to address a more geophysically relevant situation, 
the nonlinear properties of convection are then investigated in the case
when both driving agencies are destabilizing and produce sufficiently
vigorous flow. It is proposed that a meaningful approach for direct
comparison of finite-amplitude double-diffusive
convection and the better studied single-diffusive case is to compare
flows with equally large kinetic energies. Using this criterion the
characteristics of a purely thermal case, a purely compositional case
and a mixed driving case are compared. As similar flow
amplitudes purely compositional and double-diffusive cases are
characterized by a stronger time dependence compared to purely
thermal cases, and by prograde mean zonal flow near the inner
cylindrical surface. It is argued that double-diffusive cases may be
understood on the basis of purely driven ones.

Although, its low amplitude is likely to prevent double-diffusive
convection at values of the Rayleigh number significantly lower than
those for single-diffusive convection from generating magnetic fields
in the bulk of planetary cores, it is tempting to speculate that this
type of flow may have important effects in stratified layers located
just under the core-mantle boundary.  
Several mechanisms have been suggested for the for the possible
formation such layers, including the build-up of light elements
released during inner core solidification \citep{Braginsky2006},
and thermal \reva{q1}{or chemical} interaction between the mantle and the core
\citep{FearnLoper1981,Lister1995,Buffett2010}. \reva{q1}{Crucially,
evidence for stratification has been recently reported in seismic
observations of the outer core \citep{Helffrich2010}.}
Models of inert stably stratified outer layers have been found to produce
magnetic fields with morphology rather dissimilar to that of the
observed field because of a thermal wind that produces unfavorable
zonal flows throughout the core \citep{Stanley2008}. 
Inert layers, have also been found to behave like a no-slip virtual
boundary for the convective motion underneath \citep{Takehiro2010}. 
This last finding imposes a significant constraint on the flow, as it is
well known that convection structures and the morphology of the
magnetic field crucially depend on the boundary conditions
\citep{Simitev2005,Kutzner2002From,Sakuraba2009}\reva{q15}{}. The
situation may be significantly different if the stratified layer is
convecting (even weakly) rather  than inert and the low-Rayleigh
number regime \Done found here offers one such possibility. This
possibility will be subject of future research. In addition, it will
be of interest to investigate whether the results reported in this
paper hold in the more realistic 
case of a spherical shell. In particular, the spherical case may allow
the \A modes to grow to a much larger amplitude, because geostrophy is
not hard-wired into the formulation of the spherical model as it is in
the annulus case. If this should be the case, low Rayleigh-number
convection may have a more significant role in core dynamics. 
The influence of imposed magnetic fields and the general parameter
dependences of the problem must also be studied in more detail to
explore scaling relationships and the possibility of further
interesting dynamics.   

\section*{Acknowledgements}
Discussions with Prof.~F.H.~Busse, \rev{and the suggestions of an
anonymous referee} are gratefully acknowledged, as is the support of
the Royal Society under Research Grant 2010 R2.

\end{document}